\documentclass[a4paper,aip,pop,reprint,superscriptaddress]{revtex4-1}

\usepackage{graphicx}
\usepackage{indentfirst}
\usepackage{amsmath}
\usepackage{units}
\usepackage{color}
\usepackage[bookmarks=true,bookmarksnumbered=false,bookmarksopen=false,
 breaklinks=false,pdfborder={0 0 0},pdfborderstyle={},backref=false,colorlinks=true]
 {hyperref}
\hypersetup{pdftitle={Charging of dust grains by kappa plasmas},
 linkcolor=blue, citecolor=blue}
\usepackage[utf8]{inputenc}
\usepackage{amsmath,amssymb,amsfonts,textcomp}

\begin{document}

\title{Collisional charging of dust particles by suprathermal particles.
I - Standard anisotropic Kappa distributions
}
\author{L. F. Ziebell}
\email{luiz.ziebell@ufrgs.br}
\author{R. Gaelzer}
\email{rudi.gaelzer@ufrgs.br}
\affiliation{Instituto de F\'{\i}sica, Universidade Federal do Rio Grande do 
Sul, Caixa Postal 15051, CEP: 91501-970, Porto Alegre, RS, Brasil}

\date{}

\begin{abstract}
We study the effect of the velocity distributions of the
plasma particles on the equilibrium charge of dust particles which suffer
collisional charging, considering different forms of both isotropic and 
anisotropic Kappa distributions for ions and electrons. 
This paper is the first of a series of two papers on this subject. Here,
we consider
two different forms of Kappa distributions which are widely used in the 
literature, and show that effects on the dust charge associated to one of these
forms are much more significant than the effects associated to the other
basic form.
The results obtained also show that changes in the electron distribution can
have effect on the dust charge which is much more significant than the effect
which can be associated to changes in the ion distribution. 

\end{abstract}

\maketitle

\section{Introduction}
\label{sec:introduction}

Plasmas are generally defined as being ionized gases which exhibit collective
behavior. As such, they are basically composed by ions and electrons, 
sometimes containing also a population of neutral atoms or molecules.
In space environments, plasmas are composed mainly by protons and electrons, 
containing also much smaller populations of heavier ions. However, in 
addition to these conventional populations, there may be also a population
of dust particles, with a number density which is usually much smaller than
the number density of electrons and protons 
\cite{Marsch06,Schwenn06,KrugerAADDGGGHHHKLLLLMMMPSSZ06,KrugerLAG07,%
Mann2010,MannMC14}.
In the
solar wind plasma, for instance, observations made around the ecliptic plane,
over a distance range covering
from the inner region of the solar system 
up to a few astronomical units (AU) of distance from the Sun, have shown the 
presence of dust particles which are very small in size, but comprise 
significant fraction of the mass associated to the solar
wind \cite{MannKBTGMLMMGL04,Mann2008,KrugerLAG07,GrunZFG85,IshimotoM98,%
Meyer-Vernetetal09}. 
The interplanetary dust may also have effects on the
space environment around Earth and on Earth's ionosphere, as extensively 
discussed in a relatively recent review \cite{MannPMPMRMCMSN11}. 
Significant interaction
between dust and the plasma around planetary bodies has also been 
observed and studied elsewhere in the Solar System, as in Saturn's ring
system \cite{WahlundAELMSAGHKJPFRP09}.
In space environments in general, it has been argued that
the presence of dust 
may play
important role in the formation and evolution of interstellar media
\cite{TsytovichIBM14}. 

The presence of dust modifies some properties of plasmas and affects 
propagation and damping or growth of waves, including
modifications in the
properties of wave modes which also occur in dustless plasmas and also the 
identification of new modes which are due to the presence of dust. 
In the literature, the number
of articles dealing with the subject of waves in dusty plasmas has increased
almost steadily from the beginning of the years 1990, up to the beginning
of the second decade of the 21th century. A few examples may be cited,
including some papers which are
representative of the early period of the increased
research in dusty plasmas and some very recent publications, which can guide
the interested reader to other references 
\cite{Shukla1992,ShuklaS1992,Rao1993a,Rao1993b,Vladimirov1994,%
CramerV97,SalimullahR99,TsytovichA1999,TsytovichdeA00,TsytovichdeA01,
pl:deJuliSZJ05,EliassonS05,El-TaibanyS05,pl:GaelzerJZ10,pl:GalvaoZGJ11,%
KourakisSH12,MerlinoHKM12,El-LabanyEEZ14,Jatenco-PereiraCR14,JenabK14,%
pl:dosSantosZG17,pl:DeToniG21,pl:DeToniGZ22b,JahromiMBMH22,Mehdipoor22,%
pl:deToniGZ24,WeiL24}.

Regarding the populations of ions and electrons in space plasmas, the 
observations have shown that they frequently have non thermal features,
being many times anisotropic and featuring power-law tail distributions
\cite{PilippMMMRS87a,PilippMMMRS87b,MarschAT04,Marsch06}. 
Ions and electrons in the plasma wind have been frequently represented by
isotropic and anisotropic velocity distributions
which are generically known as Kappa distributions 
\cite{Vasyliunas68,%
MaksimovicPL97,MaksimovicZCISLMMSLE05,Leubner04a,StverakMTMFS09,%
PierrardL10}. Isotropic Kappa distributions appearing in the literature 
are in general in two different forms, as in 
Refs. \cite{Vasyliunas68,SummersT91,MaceHellberg95} and 
\cite{Leubner02,Leubner04b}, respectively. Anisotropic distributions 
which are based on these two different forms can also be defined, and
they are useful for the description of situations which are out
of thermodynamic equilibrium and which may excite different types of waves.

The importance of the presence of dust on the dielectric properties of a 
plasma relies on the fact that the dust particles can become charged. 
The charging of the dust particles can occur by inelastic
collisions with particles of the plasma, and also by other mechanisms, as
photoelectric charging and secondary emission
\cite{KimuraM98,Ignatov09,SodhaMM09b,pl:GalvaoZ12,pl:DeToniGZ22}.
The collisional charging of course depends on interaction with the moving
plasma particles, and therefore depends on the velocity distribution
of the plasma particles. The mechanism of charging by inelastic collisions
has been discussed in many studies in the recent literature,
some assuming Maxwellian distributions for plasma particles \cite{TriggerS99}
and some assuming that the plasma particles are described by non thermal 
distributions of power-law tipe \cite{pl:GaelzerJZ10,TribecheS11,%
TribecheS12,MishraMS13,MishraM14,MishraM15,JahromiMBMH22,MasheyevaDM22,%
Liu24}. The work described in Ref.
\cite{pl:GaelzerJZ10} is dedicated to the study of the influence of the
presence of superthermal electrons on Alfv\'en waves in solar and stellar wind
plasmas, and includes a discussion on parametric dependencies of the 
collisional charging, assuming one given form of isotropic Kappa distribution
for the electron population. In Refs.  \cite{TribecheS11,%
TribecheS12,MishraMS13,MishraM14,MishraM15,MasheyevaDM22,Liu24}, the charging of
dust particles is discussed for the case of plasma populations described by 
distributions of type Kappa, but there is no comparison between the effects 
produced by different types of distributions.

In the present paper we concentrate on the analysis of the collisional charging
of dust particles, using a well-known model for the collisional cross-section,
and study the effect of different forms of standard Kappa-type
ion and electron velocity distributions, isotropic and anisotropic. 
As far as we are aware, investigations
focused in such objective are not available in the literature.
We have also started investigations on the dust charging in the case of plasmas with regularized Kappa distributions \cite{SchererFL17}, and intend to present
the results obtained in a forthcoming publication. 
The present paper is organized as follows:
In section \ref{sec:theoretical-formulation} we briefly describe the
theoretical formulation, obtaining the conditions which must be satisfied
by the dust charge at equilibrium and the average frequency of inelastic
collisions between dust particles and plasma particles, 
considering different forms of \textcolor{blue}{standard}
Kappa velocity distribution functions
for plasma particles.
In section \ref{sec:numerical-results} we present and discuss results obtained
by numerical solution of the equilibrium condition, considering several 
possibilities for the plasma particles distribution functions, and also 
discussing the effect of variation of relevant plasma parameters.
Some final remarks are presented in section \ref{sec:conclusions}.

\section{Theoretical formulation}
\label{sec:theoretical-formulation}

The equilibrium condition for the dust-plasma system is obtained by assuming
vanishing charging current over the dust particles,
\begin{equation}
\label{chargingcurrent,1}
\sum_\beta q_\beta\int d^3v\, \sigma_\beta(v,q) v f_{\beta}({\bf v}) =0,
\end{equation}
where $\sigma_\beta$ is the cross section for dust charging by
particles of species $\beta$,
\begin{displaymath}
\sigma_\beta (v,q)=\pi a^2\left(1-\frac{2qq_\beta}{am v^2}\right)
\mbox{H}\left(1-\frac{2q_dq_\beta}{am v^2}\right),
\end{displaymath}
with $\mbox{H}$ being the Heaviside function, or {\it step} function. 
The dust particles are assumed to be spherical, with radius 
$a$, $q_d$ is the charge of a dust particle, and $q_\beta$, $m_\beta$ 
are the charge and mass of particles of species $\beta$, respectively.

Using the definition of the average collision frequency for collisional 
charging,
\begin{equation}
\label{chargingcurrent,2}
\nu_\beta= \frac{n_{d0}}{n_{\beta 0}}\int d^3v\,\frac{\pi a^2}{v}
\left(v^2-\frac{2qq_\beta}{am}\right)
\mbox{H}\left(v^2-\frac{2q_dq_\beta}{am}\right)f_\beta({\bf v}),
\end{equation}
the null current condition (\ref{chargingcurrent,1}) 
can be written as $\displaystyle
\sum_\beta q_\beta n_{\beta 0}\nu_\beta= 0$.

Moreover, we take into account the charge neutrality condition,
\begin{equation}
\label{chargeneutrality}
Ze n_{i0}-en_{e0}-Z_den_{d 0}=0,
\end{equation}
where we have written the dust charge in terms of the charge of an
electron, $q_d=-Z_ee$. With this notation, the dust charge number
$Z_d$ will be positive in the case of negative dust charge.

Using the charge neutrality condition, 
we write the condition for vanishing charging current as 
follows,
\begin{equation}
\label{chargingcurrent,3}
	Z\nu_i-\left(Z-Z_d\frac{n_{d0}}{n_{i0}}\right)\nu_e= 0
\end{equation}

The set of equations which has been presented up to this
point represents the basic formalism to be utilized for the description of the
collisional charging of dust particles. These equations can be found in many 
published sources, but nevertheless are reproduced here to give
support to the ensuing numerical analysis, which is the focus of the present
paper.

Let us introduce different forms of particle velocity distributions,
to be used in the numerical analysis. One of these is the isotropic
Kappa distribution characterized by parameters
$\alpha$ and $w_{\beta\kappa}$ \cite{pl:ZiebellG17,pl:ZiebellG19}
\begin{eqnarray}
\label{fkappa}
f_{\beta,\kappa}({\bf v})= \frac{n_{\beta 0}}
{\pi^{3/2}\kappa^{3/2} w_{\beta\kappa}^3}
\frac{\Gamma(\kappa+\alpha)}{\Gamma(\kappa+\alpha-3/2)}\\
\times
\left(1+\frac{v^2}{\kappa w_{\beta\kappa}^2}\right)^{-(\kappa+\alpha)},
\nonumber
\end{eqnarray}
where $w_{\beta\kappa}$ is a parameter with the same physical dimension as
the particle thermal velocity, which reduces to the thermal velocity in the 
limit $\kappa\to\infty$.

If the parameter $\alpha$ is taken as $\alpha=1$ and
$w_{\beta\kappa}^2= [(\kappa-3/2)/\kappa]v_\beta^2$,
with $v_\beta=\sqrt{2T_{\beta}/m_\beta}$ being the thermal velocity,
the distribution function \eqref{fkappa} becomes the 
Kappa distribution as defined in the paper by 
Summers and Thorne, 1991 
\cite{Vasyliunas68,SummersT91,MaceHellberg95}. It is also the same as
distribution denoted as type A in Ref. \cite{LazarFY16}. 
We will call this distribution
as the Kappa distribution of type I, or KI distribution. 
At this point, it is 
useful to remark that in the definition of $v_\beta$ the temperature $T_\beta$
has been written using energy units. 

If we take $\alpha=0$ and
$w_{\beta\kappa}^2= v_\beta^2$,
the distribution function \eqref{fkappa} becomes the
Kappa distribution defined in the paper by Leubner, 2002
\cite{Leubner02,Leubner04b}, which we will identify as the Kappa
distribution of type II, or KII distribution. This form of distribution
is the same as type B in Ref. \cite{LazarFY16}.

Qualitatively, the difference between KI and KII distributions is that
KI distributions have increased population of particles at small velocities,
while KII distributions have increased population at the high velocity
tail. As discussed in Ref. \cite{LazarFY16}, different physical mechanisms
can be responsible for the formation of distributions of type I and II.
These mechanisms of formation are not relevant for the present paper,
where the interest is on the discussion of the charging of dust particles by 
populations of plasma particles with different types of Kappa distributions.

Kappa distributions, either of type I or type II, may have anisotropic
temperatures. The anisotropic Kappa distribution with isotropic Kappa indexes, 
which is known as {\it bi-Kappa distribution} (BK), may be defined as follows, 
\cite{pl:ZiebellG17,pl:ZiebellG19}
\begin{eqnarray}
f_{\beta,\kappa}({\bf v})= \frac{n_{\beta 0}}
{\pi^{3/2} \kappa_\beta^{3/2} 
w_{\beta\kappa\perp}^2 w_{\beta\kappa\parallel}}
\frac{\Gamma(\kappa_\beta+\alpha)}
{\Gamma(\kappa_\beta+\alpha-3/2)}\\
\times
\left(1
+\frac{v_\parallel^2}{\kappa_\beta w_{\beta\kappa\parallel}^2}
+\frac{v_\perp^2}{\kappa_\beta w_{\beta\kappa\perp}^2}
\right)^{-(\kappa_\beta+\alpha)}, \nonumber
\label{fkappa,BK}
\end{eqnarray}

As in the case of isotropic Kappa distributions,
if the parameter $\alpha$ is taken as $\alpha=1$ and
$w_{\beta,\kappa,\parallel}^2= [(\kappa-3/2)/\kappa]v_{\beta\parallel}^2$
and $w_{\beta\kappa\perp}^2= [(\kappa-3/2)/\kappa]v_{\beta\perp}^2$, 
we obtain the bi-Kappa
distribution of type I, or BKI distribution. The bi-Kappa distribution of
type II, or BKII distribution, is obtained by assuming $\alpha=0$ and
$w_{\beta\kappa\parallel}^2= v_{\beta\parallel}^2$,
$w_{\beta\kappa\perp}^2= v_{\beta\perp}^2$.

Anisotropic Kappa distributions with anisotropic kappa indexes can also
be defined. These are known as {\it product-bi-Kappa distributions} (PBK),
and may be written as follows
\cite{pl:ZiebellG17,pl:ZiebellG19}
\begin{equation}
\begin{split}
	f_{\beta,\kappa}&({\bf v})= \frac{n_{\beta 0}}
{\pi^{3/2}\kappa_\perp \kappa_\parallel^{1/2} 
w_{\beta\kappa\perp}^2 w_{\beta\kappa\parallel}}\\
&\times\frac{\Gamma(\kappa_\perp+\alpha)\Gamma(\kappa_\parallel+\alpha)}
{\Gamma(\kappa_\perp+\alpha-1)\Gamma(\kappa_\parallel+\alpha-1/2)}\nonumber\\
&\times \left(1+\frac{v_\parallel^2}
{\kappa_\parallel w_{\beta\kappa\parallel}^2}
\right)^{-(\kappa_\parallel+\alpha)}
\left(1+\frac{v_\perp^2}
{\kappa_\perp w_{\beta\kappa\perp}^2}\right)^{-(\kappa_\perp+\alpha)}.
\nonumber
\label{fkappa,PBK}
\end{split}
\end{equation}

If the parameter $\alpha$ is taken as $\alpha=1$ and
$w_{\beta\kappa\parallel}^2= [(\kappa_\parallel-3/2)/\kappa_\parallel]
v_{\beta\parallel}^2$, 
$w_{\beta\kappa\perp}^2= [(\kappa_\perp-3/2)/\kappa_\perp]
v_{\beta\perp}^2$, 
one obtains the product-bi-Kappa
distribution of type I, or PBKI distribution. The product-bi-Kappa 
distribution of type II, or PBKII distribution, is obtained by assuming 
$\alpha=0$ and $w_{\beta\kappa\parallel}^2= v_{\beta\parallel}^2$,
$w_{\beta\kappa\perp}^2= v_{\beta\perp}^2$.

The limiting case of Kappa distribution, for large value of the parameter
kappa, is the Maxwellian distribution. For anisotropic temperatures, a 
bi-Maxwellian distribution is written as follows
\begin{equation}
f_{\beta 0}({\bf v})= \frac{n_{\beta 0}}{\pi^{3/2}v_{\beta\perp}^2
v_{\beta\parallel}}e^{-v_\parallel^2/v_{\beta\parallel}^2}
e^{-v_\perp^2/v_{\beta\perp}^2}.
\label{fbimax}
\end{equation}

Taking into account the velocity distributions functions, we can obtain the
average frequencies of inelastic collisions, which are necessary for the
determination of the equilibrium dust charge, using equation 
\ref{chargingcurrent,3}. We start by considering 
the distribution given by equation \eqref{fkappa}, and obtain
\begin{equation}
\label{nubetak,av,fkappa,3}
\begin{split}
	\left.\nu_\beta\right|_{K}
&= \frac{2(2\pi)(\pi a^{2})(\epsilon n_{i0}) v_A }
	{\pi^{3/2}\kappa_\beta^{3/2} u_{\beta\kappa}^3}
\frac{\Gamma(\kappa_\beta+\alpha_\beta)}
{\Gamma(\kappa_\beta+\alpha_\beta-3/2)}\\
&\times \frac{\kappa_\beta u_{\beta\kappa}^2}{2(\kappa_\beta+\alpha_\beta-1)}
\left(1+\frac{(u_{lim}^\beta)^2}{\kappa_{\beta}u_{\beta\kappa}^2}
\right)^{-(\kappa_\beta+\alpha_\beta-1)}\\
&\times
\left[ \frac{(\kappa_\beta+\alpha_\beta-1)\kappa_\beta u_{\beta\kappa}^2}
{\kappa_\beta+\alpha_\beta-2}
\left(1+\frac{(u_{lim}^\beta)^2}{\kappa_{\beta}u_{\beta\kappa}^2}
\right)\right.\\
&\left.
+\frac{2Z_dZ_\beta e^2}{a m_\beta v_A^2}\right],
\end{split}
\end{equation}
where $Z_i$ indicates the ion charge number, $Z_e=-1$, 
$\epsilon$ is the number density of dust particles divided by the 
equilibrium ion density, $\epsilon=n_{d0}/n_{i0}$, $v_A$ is the
Alfv\'en velocity, $v_A=B_0/\sqrt{4\pi n_{i0}m_i}$, and
\begin{displaymath}
u_{lim}^e=\left(\frac{2Z_{d}e^2}{a m_e v_A^2}
\right)^{1/2}, \,\,\,\,\,\,\,\,\,\,
u_{lim}^i= 0~. 
\end{displaymath}

For a bi-Kappa distribution function, \eqref{fkappa,BK}, we 
obtain
the following,
\begin{equation}
\label{nubetak,av,BK,4}
\begin{split}
&\left.\nu_\beta\right|_{BK}=  
\frac{2(2\pi)(\pi a^{2})(\epsilon n_{i0}) v_A }{\pi^{3/2} \kappa_\beta^{3/2} 
u_{\beta\kappa\perp}^2 u_{\beta\kappa\parallel}}\\
&\times\frac{\Gamma(\kappa_\beta+\alpha_\beta)}
{\Gamma(\kappa_\beta+\alpha_\beta-3/2)} 
\frac{(\kappa_{\beta}u_{\beta\kappa\perp}^2)^{\kappa_\beta+a_\alpha}}{2}
\\
&\times\int_0^1 d\mu 
\frac{1}{\left[1-\mu^2(1-u_{\beta\kappa\perp}^2/u_{\beta\kappa\parallel}^2)
\right]
(\kappa_\beta+a_\beta-1)} \\
& \times
\left[\kappa_\beta u_{\beta\kappa\perp}^2 + (u_{lim}^\beta)^2
\left[1-\mu^2(1-u_{\beta\kappa\perp}^2/u_{\beta\kappa\parallel}^2)\right]
\right]^{-(\kappa_\beta+a_\beta-1)}\\
&\times
\left[\frac{\kappa_\beta u_{\beta\kappa\perp}^2
+ (u_{lim}^\beta)^2
\left[1-\mu^2(1-u_{\beta\kappa\perp}^2/u_{\beta\kappa\parallel}^2)\right]
(\kappa_\beta+a_\beta-1)}
{\left[1-\mu^2(1-u_{\beta\kappa\perp}^2/u_{\beta\kappa\parallel}^2)\right]
(\kappa_\beta+a_\beta -2)}\right.\\
&\left.
+\frac{2Z_dZ_\beta e^2}{a m_\beta v_A^2}\right],
\end{split}
\end{equation}
with $u_{\beta\kappa,\parallel}= w_{\beta\kappa\parallel}/v_A$ and
$u_{\beta\kappa,\perp}= w_{\beta\kappa\perp}/v_A$.

For a PBK distribution \eqref{fkappa,PBK}, the average collision frequency
can be obtained as follows,
\begin{equation}
\label{nubetak,av,PBK,2}
\begin{split}
&\left.\nu_\beta\right|_{PBK}=  
\frac{2(2\pi)(\pi a^{2})(\epsilon n_{i0}) v_A}
{\pi^{3/2} \kappa_{\beta\perp}\kappa_{\beta\parallel}^{1/2} 
u_{\beta\kappa\perp}^2 u_{\beta\kappa\parallel}}\\
&\times\frac{\Gamma(\kappa_{\beta\perp}+\alpha_\beta)
\Gamma(\kappa_{\beta\parallel}+\alpha_\beta)}
{\Gamma(\kappa_{\beta\perp}+\alpha_\beta-1)
\Gamma(\kappa_{\beta\parallel}+\alpha_\beta-1/2)}
\\
&\times 
\int_{x_{min}^\beta}^{\pi/2} dx~\frac{\tan(x)}{\cos^2(x)} 
\left(\tan^{2}(x)
+\frac{2Z_dZ_\beta e^2}{a m_\beta v_A^2}
\right)
\\
& \times
\int_{0}^1 d\mu
\left(1+\frac{\tan^2(x)\mu^2}
{\kappa_{\beta\parallel} u_{\beta\kappa\parallel}^2}
\right)^{-(\kappa_{\beta\parallel}+\alpha_\beta)}\\
&\times
\left(1+\frac{\tan^2(x) (1-\mu^2)}
{\kappa_{\beta\perp} u_{\beta\kappa\perp}^2}
\right)^{-(\kappa_{\beta\perp}+\alpha_\beta)}~,
\end{split}
\end{equation}
where we have defined $u=\tan(x)$ and where
$x_{min}^\beta= \tan^{-1}(u_{lim}^\beta)$,
with $u_{\beta\kappa,\parallel}= w_{\beta\kappa\parallel}/v_A$ and
$u_{\beta\kappa,\perp}= w_{\beta\kappa\perp}/v_A$.

Using equation \eqref{fbimax}, the Maxwellian distribution 
function, we obtain the following form for
the average collisional frequency,
\begin{equation}
\begin{split}
\label{nubetak,av,bimax}
\nu_\beta
&= (\pi a^{2}) (\epsilon n_{i0}) v_A 
\frac{2\pi}{\pi^{3/2}}\frac{u_{\beta\parallel}}{\Delta_\beta} \\
&\times \int_{0}^1 d\mu~ 
\frac{e^{-t_{lim}^\beta[1-\mu^2(1-\Delta_\beta)]/\Delta_\beta}}
{[1-\mu^2(1-\Delta_\beta)]^2/\Delta_\beta^2}\\
&\times 
\left[1+\delta_{\beta i} \frac{[1-\mu^2(1-\Delta_\beta)]}{\Delta_\beta}
\frac{2Z_{d}^kZ_{i}e^2}{a m_i v_A^2 u_{i\parallel}^2}
\right],
\end{split}
\end{equation}
where $t_{lim}^\beta= (u_{lim}^\beta)^2/u_{\beta\parallel}^2$.

\section{Numerical Analysis}
\label{sec:numerical-results}

For the numerical analysis, we consider plasma parameters which may be
considered representative of plasmas in the solar wind environment. The
exact value of these parameters is not relevant in the present context,
where the emphasis is on the effect of the shape of the velocity distribution
functions on the collisional charging of dust particles.
We therefore consider as example $a=\unit[1.0\times 10^{-4}]{cm}$ and
$\epsilon= 1.0\times 10^{-5}$. We also consider $\beta_i=2.0$, 
$v_A/c=1.0\times 10^{-4}$, and $n_{i0}=\unit[10]{cm^{-3}}$, and assume that 
$T_{e\parallel}=T_{i\parallel}$, and
that the ions are protons, so that $Z_i=1$.
Moreover, from the value
of $\beta_i$ and $v_A/c$, taking into account the mass of the ions, we
can obtain $T_{i\parallel}$. 

As the starting point of the numerical analysis, we assume that
ions and electrons are described by isotropic Maxwellian distributions, 
solve equation \eqref{chargingcurrent,3}, and obtain the equilibrium value
of the charge number in a dust particle, $Z_d=15{,}470$. If the electrons
are described by an isotropic electron distribution and the ions are described 
by an anisotropic Maxwellian distribution with 
$T_{i\perp}/T_{i\parallel}=0.2$, we obtain $Z_d=14{,}182$, and if 
$T_{i\perp}/T_{i\parallel}=5.0$, $Z_d=15{,}749$. On the other hand, if
the velocity distribution of the ions is an isotropic Maxwellian, and the
electrons are described by an anisotropic Maxwellian with
$T_{e\perp}/T_{e\parallel}=0.2$, we obtain $Z_d=8{,}774$, and
$Z_d=44{,}182$ for the case  
$T_{e\perp}/T_{e\parallel}=5.0$. These results obtained considering isotropic
and anisotropic Maxwellian distributions appear listed in Table
\ref{tab1}. General conclusions to be drawn from these results are that the
dust charge increases with the increase of the ratio $T_\perp/T_\parallel$,
and that the dust charge is strongly affected by changes in the temperature
anisotropy in the electron distribution and is only slightly affected by
the anisotropy in the ion distribution.
\begin{table}[h]
\begin{tabular}{|l|c|} \hline
Type of distribution \quad\quad & \quad Dust charge number \quad \\
\hline\hline
e Max, i Max & 15{,}470 \\ \hline
e bi-Max, $T_{e\perp}/T_{e\parallel}=0.2$, i Max & 8{,}774 \\ \hline
e bi-Max, $T_{e\perp}/T_{e\parallel}=5.0$, i Max & 44{,}382 \\ \hline
e Max, i bi-Max, $T_{i\perp}/T_{i\parallel}=0.2$ & 14{,}182 \\ \hline
e Max, i bi-Max, $T_{i\perp}/T_{i\parallel}=5.0$ & 15{,}749 \\ \hline
\hline
\multicolumn{2}{c}{}
\end{tabular}
\caption{
Charge number on spherical dust particles collisionally charged, for
different cases of isotropic and anisotropic Maxwellian distribution for 
electrons and ions.
$\beta_i=\unit[2.0]{}$, $v_A/c=\unit[1.0\times 10^{-4}]{}$, 
$a=\unit[1.0\times 10^{-4}]{cm^{-3}}$, 
$\epsilon=\unit[1.0\times 10^{-5}]{}$,
$n_{i0}=\unit[1.0\times 10]{cm^{-3}}$.
}
\label{tab1}
\end{table}

\begin{figure}
\hspace{-0.52cm} \includegraphics[scale=0.6]{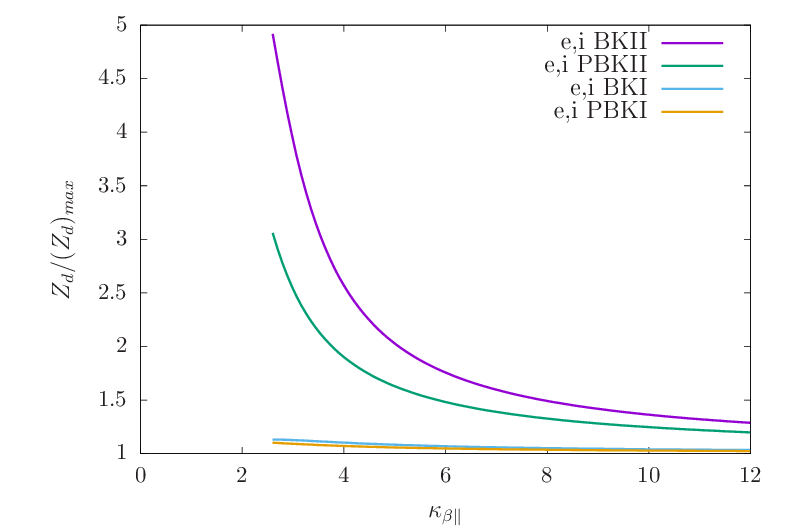}
\caption{
Relative charge number on spherical dust particles collisionally charged, 
for isotropic temperatures and different forms of Kappa distributions. In
each case depicted, ions and electrons have the same form of velocity
distribution.
The parameters are $\beta_i=2.0$, $v_A/c= 1\times 10^{-4}$, 
$\epsilon= 1.0\times 10^{-7}$, $a=\unit[1.0\times 10^{-4}]{cm}$, with
$T_{e\parallel}=T_{i\parallel}$.
}
\label{fig1}
\end{figure}

The effect of the change of the kappa index, in the case of Kappa velocity
distributions, can be investigated considering different situations.
Initially, we consider that ions and electrons have the same type of 
nonthermal distribution, with isotropic temperatures (i.e., $T_{\beta\perp}
=T_{\beta\parallel}$). Figure \ref{fig1} 
illustrates
the equilibrium dust charge
obtained from equation \ref{chargingcurrent,3} in this case, 
by considering different forms of
the velocity distribution.
Figure \ref{fig1}, as well as all the ensuing figures, shows the values of
the relative dust charge, which is the equilibrium dust charge
divided by the value obtained in the case of isotropic Maxwellian 
distributions for ions and electrons (i.e., $Z_d=(Z_d)_{max}= 15{,}470$, 
according to table \ref{tab1}). The values are shown as a function of
$\kappa_{\beta\parallel}$, the kappa index of the velocity distribution along 
the parallel direction. Of course, for BKI and BKII distributions the
values of the parallel and perpendicular kappa indexes are the same,
$\kappa_{\beta\parallel}=\kappa_{\beta\perp}=\kappa_\beta$. 
For the PBK distributions,
we consider $\kappa_{\beta\perp}=\kappa_{\beta\parallel}$. 

The two lower
curves appearing in figure \ref{fig1} show that for distributions of type I,
either BKI or PBKI, the relative dust charge is very close to 1. The most
significant effect occurs for $\kappa_{\beta}$ close to 2.5, and is at most
about 15\%, for BKI distribution. For type II distributions the change is much
more significant. It is seen that for small values of $\kappa_{\beta\parallel}$
the dust charge is increased by nearly 400\% in the case of BKII distributions,
as compared to the case of Maxwellian distributions,
and 200\% in the case of PBKII distributions.

\begin{figure*}
\hspace{-0.53cm} \includegraphics[scale=1.2]{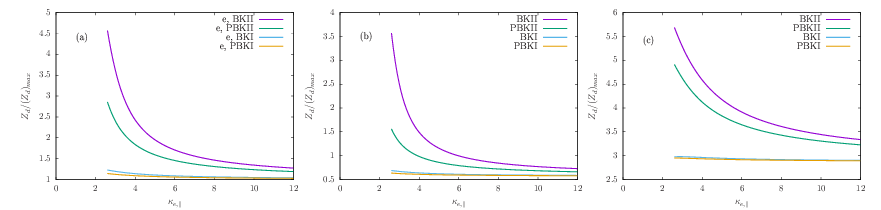}
\caption{
Relative charge number on spherical dust particles collisionally charged, in the case
of isotropic Maxwellian distribution for ions and different forms of 
Kappa distributions for electrons, isotropic and anisotropic.
The figure shows the values of the charge
number divided by the charge number obtained in the case of isotropic
Maxwellian distribution for ions and electrons. 
(a) $T_{e\perp}/T_{e\parallel}=1.0$;
(b) $T_{e\perp}/T_{e\parallel}=0.2$;
(c) $T_{e\perp}/T_{e\parallel}=5.0$;
Parameters are as in figure \protect{\ref{fig1}}.
}
\label{fig2}
\end{figure*}

In figure \ref{fig2} we consider cases in which the electrons are described
by a BK  distribution or a PBK distribution, while the ions are described by an 
isotropic Maxwellian distribution. Figure \ref{fig2}(a) depicts the
case of isotropic value of $T_e$, figure \ref{fig2}(b) shows the results
obtained with $T_{e\perp}/T_{e\parallel}=0.2$, and figure \ref{fig2}(c)
the results obtained with $T_{e\perp}/T_{e\parallel}=5.0$.
The three panels of figure \ref{fig1}
show that in the case of electron distributions of type I the dust charge is
almost the same as in the Maxwellian case, with the charge slightly
larger in the case of BKI distribution than in the case of PBKI distributions. 
The comparison between figure
\ref{fig2}(b) and figure \ref{fig2}(a) 
show that for distributions of type I the dust charge
decreases by a factor of nearly two when the electron temperature ratio is 
changed from 1 to 0.2, in the region of small $\kappa_{e\parallel}$. 
The comparison
between figure \ref{fig2}(c) and figure \ref{fig2}(a) shows that for type I 
distributions the dust charge is increased in the region of small 
$\kappa_{e\parallel}$ by a factor of nearly 2.5, 
when the electron temperature ratio is changed from 
1 to 5. For the same changes of electron temperature ratio, table \ref{tab1}
shows that in the bi-Maxwellian case the dust charge changes by factors of 
0.57 and 2.87, respectively.
It is therefore seen that regarding the charging of dust particles the 
situation in the case of a plasma with electrons described by type I 
distributions, either BKI or PBKI, is not significantly
different from the situation in the bi-Maxwellian case.
On the other hand, in the case of distributions of type II, either BKII or 
PBKII, figures \ref{fig2}(a,b,c) show 
that the charge of the dust particles is always above that 
obtained for type I distributions, for the whole range of values of
$\kappa_{e\parallel}$ which has been considered, with steep increase for
small values of $\kappa_{e\parallel}$. 
In the whole interval of electron temperature ratio, from 
$T_{e\perp}/T_{e\parallel}=0.2$ to 5, the dust charge is greater
in the case of BKII distribution than in the case of PBKII distribution. The
difference between PBKII and BKII distributions is more noticeable for
temperature ratio smaller than 1 than for temperature ratio greater than 1.
For instance, in the case of electron temperature ratio 0.2, shown in the 
figure \ref{fig2}(b), the dust charge for $\kappa_{e\parallel}=
2.5$ and BKII distribution is more than twice the dust charge obtained for
PBKII distribution. For temperature ratio 5, shown in figure \ref{fig2}(c), the 
dust charge at the same value of $\kappa_{e\parallel}$ and BKII is only about
1.2 times the value obtained with a PBKII distribution.

\begin{figure*}
\hspace{-0.53cm} \includegraphics[scale=1.2]{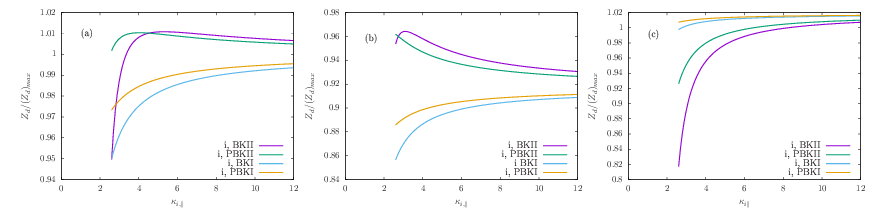}
\caption{
Relative charge number on spherical dust particles collisionally charged, in the case
of isotropic Maxwellian distribution for electrons and different forms of 
Kappa distributions for ions, isotropic and anisotropic.
The figure shows the values of the charge
number divided by the charge number obtained in the case of isotropic
Maxwellian distribution for ions and electrons. 
(a) $T_{i\perp}/T_{i\parallel}=1.0$;
(b) $T_{i\perp}/T_{i\parallel}=0.2$;
(c) $T_{i\perp}/T_{i\parallel}=5.0$;
Parameters are as in figure \protect{\ref{fig1}}.
}
\label{fig3}
\end{figure*}

In figure \ref{fig3} we change the roles of the particle distribution function,
in comparison with figure \ref{fig2}. That is, in figure \ref{fig3}
we consider cases in which the ions are described by a BK 
distribution or a PBK distribution, and the electrons are described by an 
isotropic Maxwellian distribution. 
Panel (a) of figure \ref{fig3} shows cases with 
$T_{i\perp}/T_{i\parallel}= 1.0$, panel (b) shows cases with 
$T_{i\perp}/T_{i\parallel}= 0.2$, and panel (c) shows cases with
$T_{i\perp}/T_{i\parallel}= 5.0$. In all these three panels, we show results
obtained considering four different types of the ion distribution, BKI, BKII,
PBKI, and PBKII. For distributions of type I, BKI and PBKI, it is seen in
figure \ref{fig3}(a) that
in the case of isotropy of temperatures the dust charge is smaller than in the
case of Maxwellian ion distribution, by a small amount, with the maximum 
difference occurring for small value of $\kappa_{i\parallel}$. 
Figure \ref{fig3}(b)
shows that the effect is similar for the case of $T_{i\perp}<T_{i\parallel}$,
with increase of the difference relative to the case of isotropic Maxwellian. 
For $T_{i\perp}>T_{i\parallel}$, as illustrated by figure \ref{fig3}(c), 
it is seen
that the normalized dust charge is slightly above 1 for the whole range of
values of $\kappa_i$, except for a small decrease for the case of PBKI
distribution, near $\kappa_i=2.5$.
For the case of BKII and PBKII ion distributions the difference relative
to the isotropic Maxwellian case is more noticeable, but still small. Figures
\ref{fig3}(a,b,c) show that is that in the case of distributions of type II 
the dust 
charge ratio tends to
increase for decreasing values of $\kappa_{i\parallel}$, for
$T_{i\perp}<T_{i\parallel}$ (figure \ref{fig3}(b)), with inversion of this
behavior
for $\kappa_{i\parallel}$ approaching 2.5, in the case of BKII distribution.
For $T_{i\perp}=T_{i\parallel}$, figure \ref{fig3}(a)
shows nearly the same tendency,
with the difference that both in the case of BKII and PBKII ion distributions
the dust charge ratio decreases for very small $k_{i\parallel}$.
For $T_{i\perp}>T_{i\parallel}$, according to the results shown in 
figure \ref{fig3}(c),
the dust charge ratio is smaller than unity for small
$k_{i\parallel}$, and tends uniformly to unity for increasing values of
$\kappa_{i\parallel}$. The effect is more pronounced for BKII distribution
than for PBKII distribution.

\begin{figure*}
\hspace{-0.42cm} \includegraphics[scale=1.2]{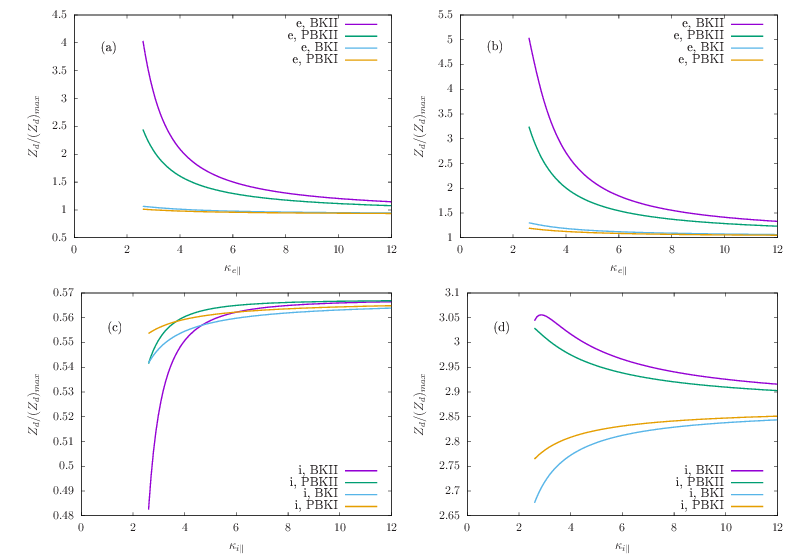}
\caption{
Relative charge number on spherical dust particles collisionally charged.
The figure shows the values of the charge
number divided by the charge number obtained in the case of isotropic
Maxwellian distribution for ions and electrons. 
(a) Bi-Maxwellian distribution for ions, with $T_{i\perp}/T_{i\parallel}=0.2$,
and different forms of Kappa distributions for electrons, with 
$T_{e\perp}/T_{e\parallel}=1.0$;
(b) Bi-Maxwellian distribution for ions, with $T_{i\perp}/T_{i\parallel}=5.0$,
and different forms of Kappa distributions for electrons, with 
$T_{e\perp}/T_{e\parallel}=1.0$;
(c) Bi-Maxwellian distribution for electrons, with
$T_{e\perp}/T_{e\parallel}=0.2$,
and different forms of Kappa distributions for ions, with 
$T_{i\perp}/T_{i\parallel}=1.0$;
(d) Bi-Maxwellian distribution for electrons, with
$T_{e\perp}/T_{e\parallel}=5.0$,
and different forms of Kappa distributions for ions, with 
$T_{i\perp}/T_{i\parallel}=1.0$;
Parameters are as in figure \protect{\ref{fig1}}.
}
\label{fig4}
\end{figure*}

In figure \ref{fig4} we present results obtained considering that one of the
plasma species has a bi-Maxwellian distribution, and the other species is
described by Kappa distributions with isotropy of temperatures. Figure
\ref{fig4}(a) displays results for the case in which the ions are described
by a bi-Maxwellian distribution with $T_{i\perp}/T_{i\parallel}=0.2$, and
the electrons feature BKI, PBKI, BKII or PBKII distributions, with 
$T_{e\perp}=T_{e\parallel}$.
Figure
\ref{fig4}(b) displays results for the case in which the ions are described
by a bi-Maxwellian distribution with $T_{i\perp}/T_{i\parallel}=5.0$, and
the electrons feature BKI, PBKI, BKII or PBKII distributions, with 
$T_{e\perp}=T_{e\parallel}$.
For panels (c) and (d) the configurations is reversed. Figure 
\ref{fig4}(c) shows results for the case in which the electrons have
a bi-Maxwellian distribution with $T_{e\perp}/T_{e\parallel}=0.2$, and
the ions have BKI, PBKI, BKII or PBKII distributions, with 
$T_{i\perp}=T_{i\parallel}$, and figure
\ref{fig4}(d) exhibits results for electrons described
by a bi-Maxwellian distribution with $T_{e\perp}/T_{e\parallel}=5.0$, and
the ions have BKI, PBKI, BKII or PBKII distributions, with 
$T_{i\perp}=T_{i\parallel}$.
Panels (a) and (b) of figure \ref{fig4} show that in the case of ions with
bi-Maxwellian distribution the anisotropy of ion temperatures is not very
significant for the charging of the dust particles. 
Indeed, the results obtained 
in panels (a) and (b) are very similar to those appearing in figure 
\ref{fig2}(a), in which the ion distribution was isotropic. It is only noticed
that for $T_{i\perp}>T_{i\parallel}$ the magnitude of the negative charge
of dust particles is higher than in the isotropic case, and for
$T_{i\perp}<T_{i\parallel}$ it is smaller than in the isotropic case.
On the other hand, figures \ref{fig4}(c) and \ref{fig4}(d) show that the
anisotropy in a electron Maxwellian distribution has much more significant
effects on the dust charge. The comparison between figure \ref{fig4}(c) and
figure \ref{fig2}(a) show that for $T_{e\perp}<T_{e\parallel}$ the magnitude
of the dust charge tends to be reduced in comparison with the isotropic case,
for all types of Kappa distributions considered for the ions. In the case of
$T_{e\perp}=0.2T_{e\parallel}$ considered in figure \ref{fig4}(c), the 
magnitude of the dust charge is nearly 1/2 of the value in the isotropic case,
for most of the range of values of $\kappa_{i\parallel}$. As in the isotropic
case, there is a reduction in the magnitude of the charge for small values
of $\kappa_{i\parallel}$, which is most pronounced in the case of ion BKII
distribution. Conversely, figure \ref{fig4}(d) shows that in the case of 
$T_{e\perp}>T_{e\parallel}$ the magnitude of the dust charge increases in
comparison with the value obtained for electron isotropic Maxwellian
distribution, for all forms of ion Kappa distribution which were considered.
The most pronounced effect occurs for ion BKII distribution.

\begin{figure*}
\hspace{-0.42cm} \includegraphics[scale=1.2]{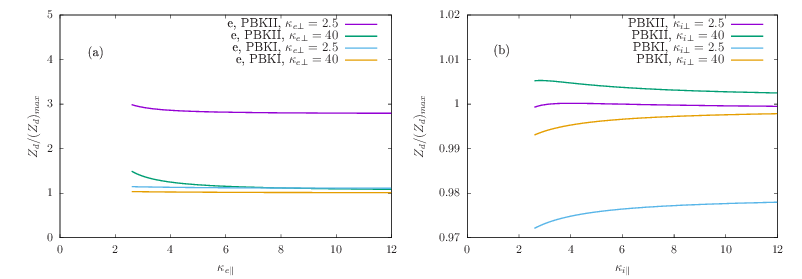}
\caption{
Relative charge number on spherical dust particles collisionally charged, vs
the parallel kappa index.
(a) Isotropic Maxwellian distribution for ions, and PBK distribution for
electrons, with isotropic temperatures and two values of $\kappa_{e\perp}$, 
2.5 and 40;
(b) Isotropic Maxwellian distribution for electrons, and PBK distribution for 
ions, with isotropic temperatures and two values of $\kappa_{i\perp}$, 2.5 and
40.
Parameters are as in figure \protect{\ref{fig1}}.
}
\label{fig5}
\end{figure*}

We have also investigated the
effect of anisotropy of the $\kappa$ parameters in PBK distributions,
presenting some results in figure \ref{fig5}. For this figure, we have 
considered that one of the plasma species is described by an isotropic
Maxwellian distribution, and the other by a PBK distribution with isotropic
temperature parameter and anisotropic kappa indexes. 
In the case of figure \ref{fig5}(a), we have considered that the ions have 
an isotropic Maxwellian distribution, and the electrons a PBK distribution.
The relative dust charge in shown as a function of $\kappa_{e\parallel}$, for
the cases of $\kappa_{e\perp}=2.5$ and $\kappa_{e\perp}=40$, 
for each type of PBK electron distribution.

The case with 
$\kappa_{e\perp}=2.5$ is shown in figure \ref{fig5}(a) by lines magenta and 
blue, for electron distributions of type PBKII and PBKI, respectively. 
In this case, for small $\kappa_{e\parallel}$
the electron distribution features non-thermal tails along parallel and 
perpendicular directions, and 
for increasing values of $\kappa_{e\parallel}$ 
tends to a state which is nearly Maxwellian along
parallel direction with non-thermal tail along perpendicular direction.
The blue line shows that for PBKI electron distribution 
the dust charge is slightly above the charge in the
Maxwellian case for the whole range of $\kappa_{e\parallel}$ considered in 
the figure, and nearly insensitive to this parameter. 
The magenta line shows that for PBKII electron distribution the dust charge
is nearly three times the charge obtained in the Maxwellian case, and
is almost insensitive to the change in $\kappa_{e\parallel}$, except
for a small increase near $\kappa_{e\parallel}=2.5$.

The case of $\kappa_{e\perp}=40$ is depicted in figure
\ref{fig5}(a) by lines orange and green, corresponding to PBKI and PBKII 
electron distributions, respectively. The value of $\kappa_{e\perp}$ in this
case is such that the velocity distribution is almost Maxwellian along
perpendicular direction. Accordingly, figure \ref{fig5}(a) shows that the
normalized dust charge is close to 1 for most of the range of 
$\kappa_{e\parallel}$ depicted in the figure, with
slightly larger value in the case of PBKII distribution (green line), more 
pronounced for small value of $\kappa_{e\parallel}$, where the nonthermal
feature along parallel direction becomes more noticeable.

In figure \ref{fig5}(b) we consider the case of isotropic Maxwellian 
distribution for electrons and PBK distribution for ions. The orange and
green lines correspond to the case of ion distribution PBKI and PBKII, 
respectively, with $\kappa_{i\perp}=40$. In both cases the relative dust
charge is close to 1, for all values of $\kappa_{i\parallel}$, from 2.5 up to
12. The value is slightly smaller than 1 in the case of PBKI ion distribution,
and slightly larger than 1 for PBKII ion distribuution. For $\kappa_{i\perp}
=2.5$, shown by lines blue and magenta, for PBKI and PBKII ion distributions,
respectively,
the magenta line shows that the charge is very close the value in the 
Maxwellian case, for PBKII ion distribution, and nearly 3\% below the 
Maxwellian value for PBKI ion distribution. The reduction of dust charge
value is slightly more pronounced than in the case of PBKI for
$\kappa_{i\perp}=40$ (orange line), but nevertheless not very significant.
The four cases considered appear very close in figure \ref{fig5}(b),
indicating that the anisotropy in the $\kappa$ indexes in the ion distribution
is not very relevant for the determination of the equilibrium charge of a dust
particle.

\section{Final remarks}
\label{sec:conclusions}

In the present paper we have discussed the effect of different forms of
Kappa velocity distributions of the plasma particles on the collisional 
charging of dust particles immersed in the plasma. We have used a well-known
model for the collisional charging, and used in the analysis both velocity
distributions known as bi-Kappa distributions and velocity
distributions known as product-bi-Kappa distributions. Thse distributions 
have been based on forms which were disseminated
in works like those of Vasyliunas (1968), Summers \& Thorne (1991) and
Mace \& Hellberg (1995) \cite{Vasyliunas68,SummersT91,MaceHellberg95}, and
also based on forms which can be found in papers as those by Leubner
(2002, 2004) \cite{Leubner02,Leubner04b}. Bi-Kappa distributions based on
these two basic models were denominated BKI and BKII distributions, 
respectively, and product-bi-Kappa distributions based on these two types
were denominated PBKI and PBKII, respectively. 
Using these velocity distributions, we have discussed situations with
isotropy of temperature parameters as well as situations with temperature
anisotropy, including a discussion of the effect of the variation of the
parallel and perpendicular kappa indexes of the velocity distributions.

The results obtained have shown that the influence of the different types of 
velocity distributions on the dust charging is more significant for
small values of the kappa indexes, for all types of distributions which have
been considered. For all the cases 
investigated, the charge acquired by dust
particles due to inelastic collision processes has been negative. 
We have 
not obtained situations with positive dust charges originated from collisional
processes, 
reported recently in connection with
regularized Kappa distribution functions \cite{Liu24}. However,
We are presently investigating 
the charging of dust particles by plasma particles described by these 
regularized Kappa distributions, and intend to publicize our results in a 
forthcoming publication.

The results obtained have shown that the effect of the modification of the
ion distribution function,
considering electrons with isotropic Maxwellian distributions,
leads only to a small change of the charge of dust particles. 
It has been seen that in these cases the change in the dust particle charge 
is at most around 20\%, for very small value of the kappa index, tending
to be negligible for larger values of kappa. Moreover, if the
electrons are characterized by anisotropic Maxwellian distributions, more
significant effect may be observed, but the influence of the change in the
Kappa distributions of ions remains small. The small effect of the change
of the ion distribution is perhaps not so surprising, since the charging
process associated to inelastic collisions 
is dominated by the very mobile electrons,
so that the dust charge results to be negative.

On the other hand, the results obtained have shown that the influence of the
form of the electron distribution
on the dust particle charge can be significant. The 
effect is more noticeable in the region of small values of the kappa index, and
in this region the difference 
between charges produced by distributions of type I, BKI and 
PBKI, and by distributions of type II, BKII and PBKII, 
can be very significant.
For the study of the effect
of the electron distribution, we have considered situations in which ions have
isotropic and anisotropic Maxwellian distributions. For isotropic ion distribution,
it has been seen that BKI and PBKI distributions for electrons, with isotropic
temperatures, lead to dust particles with charges which differ of the 
isotropic Maxwellian case at most by a factor which approaches 25\%, for small 
value of the parallel kappa index. For BKII and PBKII electron distributions, 
however, the effect
can be much more significant, attaining more than 300\%, for BKII 
distributions, near $\kappa_{e\parallel}=2.5$. For anisotropic temperatures in
the electron distributions,
the effect of Kappa distributions on the dust charge can be more notorious,
but continues to be more significant for type II distributions than for type
I distributions. Similar conclusions was also drawn from cases with bi-Maxwellian
distributions for the ions.

\acknowledgments
LFZ acknowledges support from CNPq (Brazil), grant No. 303189/2022-3.
RG acknowledges support from CNPq (Brazil), grant No. 313330/2021-2. 
This study was financed in part by the Coordena\c{c}\~ao de Aperfei\c{c}oamento
de Pessoal de N\'{\i}vel Superior - Brasil (CAPES) - Finance Code 001.

\begin{itemize}
\item ORCID:\\
Luiz F. Ziebell : 0000-0003-0279-0280\\
Rudi Gaelzer: 0000-0001-5851-7959
\end{itemize}

\section*{Declarations}

\begin{itemize}

\item Conflict of interest

The authors have no financial or non-financial interests to declare.

\item Authors' contributions

Both authors contributed to the conception and design of the manuscript, and
worked on the development of the research. Luiz F. Ziebell wrote the first 
draft of the manuscript, and both authors contributed to the preparation of
the version to be submitted. Both authors read and approved the final
manuscript.

\item Data availability

The data that support the findings of this study are available within the
article.

\end{itemize}


\end{document}